\documentclass{article}
\usepackage{stmaryrd}
\usepackage{mathrsfs}
\usepackage{amsmath}
\usepackage{amssymb}
\usepackage{amsfonts}
\usepackage{pstricks}
\usepackage{graphicx, color, epsfig, enumerate}
\newcommand{\pa}{\partial}

\newcommand{\ga}{\gamma}
\newcommand{\Ga}{\Gamma}

\newcommand{\la}{\lambda}

\newcommand{\al}{\alpha}
\newcommand{\be}{\beta}

\newcommand{\f}{\frac}

\begin{document}
\title{{Start-up flow of a viscoelastic fluid in a pipe}\\
{with fractional Maxwell's model\footnote{The first submission was
in July 2008. A second submission was to Computers and Mathematics
with Applications in April 2009. This is a revised edition of the
second submission.}}}
\author{
\\
{Di Yang\footnote{yangd04@mails.tsinghua.edu.cn} ~~~~~ Ke-Qin Zhu\footnote{zhukq@tsinghua.edu.cn}}\\
{\small Department of Engineering Mechanics,}\\
{\small Tsinghua University, Beijing 100084, P. R. China}\\
\date{}
}
\maketitle

\begin{abstract}
Unidirectional start-up flow of a viscoelastic fluid in a pipe with
fractional Maxwell's model is studied. The flow starting from rest
is driven by a constant pressure gradient in an infinite long
straight pipe. By employing the method of variable separations and
Heaviside operational calculus, we obtain the exact solution, from
which the flow characteristics are investigated. It is found that
the start-up motion of fractional Maxwell's fluid with parameters
$\alpha$ and $\beta$, tends to be at rest as time goes to infinity,
except the case of $\beta=1$. This observation, which also can be
predicted from the mechanics analogue of fractional Maxwell's model,
agrees with the classical work of Friedrich and it indicates
fractional Maxwell's fluid presents solid-like behavior if $\be\neq
1$ and fluid-like behavior if $\be=1$. For an arbitrary viscoelastic
model, a conjecture is proposed to give an intuitive way judging
whether it presents fluid-like or solid-like behavior. Also
oscillations may occur before the fluid tends to the asymptotic
behavior stated above, which is a common phenomenon for viscoelastic
fluids.
\end{abstract}

\section{Introduction}
`All things are movable and in a fluid state', which is a famous
quotation from Thales of Miletos, the first philosopher of ancient
Greece.

Indeed, besides the most familiar fluids such as water and gas, most
materials in nature and industry, such as milk, oil, lava, etc., can
be treated and investigated as fluids. However, some of them not
only have the viscosity like Newtonian fluid, but also exhibit
Hooke's elasticity. They are viscoelastic materials. Different
models of viscoelastic materials were obtained and studied during
the past hundreds of years; for example, Maxwell's model,
constructed by a spring and a dashpot in serial, was largely
investigated in the last century.

Recently, fractional Maxwell's model, constructed by two fractional
element models in serial, attracts a lot of researchers'
interests\cite{YZ}\cite{ZHY}\cite{Tan}\cite{Herna}\cite{Hayat}\cite{Palade}\cite{Fetecau_2}\cite{Fetecau_3}\cite{Fetecau_4}.
Let $\sigma$ be the shear stress and $\epsilon$ be the shear strain.
The constitutive equation for fractional Maxwell's model is given by
\begin{equation}\label{constitutive}
\sigma+\la^\al\f{d^\al}{dt^\al}
\sigma=E\la^\be\f{d^\be}{dt^\be}\epsilon,~~~~~0\leq\al\leq\be\leq 1,
\end{equation}
where $E$ the shear modulus, and $\la$ is the relaxation
time.\footnote{Some studies on fractional Maxwell's model only
concern the particular case with $\be=1$, i.e., the so called
generalized Maxwell's model.}

In the case of $\al=0$, equation \eqref{constitutive} degenerates to
$$\sigma=\f{1}{2}E\la^\be \f{d^\be}{dt^\be}\epsilon,$$
which is just the constitutive relation of a fractional element
model with the shear modulus $E/2$. In fact, the usual expression
for a fractional element model was first introduced by Scott
Blair\cite{G. W. Scott Blair}\cite{Scott2}
\begin{equation}
\sigma=E_s\la^\be \f{d^\be}{dt^\be}\epsilon,
\end{equation}
where $E_s$ is the shear modulus. The mechanics analogue of a
fractional element model can be found in
\cite{YZ}\cite{ZHY}\cite{Heymans}\cite{Pod}.

In the case of $\al=\be=1$, equation \eqref{constitutive}
degenerates to the constitutive relationship of classical Maxwell's
model.

Physically, fractional Maxwell's model can be considered as two
fractional element models in serial, with orders $\ga_1$ and $\ga_2$
satisfying
\begin{equation}
\al=|\ga_1-\ga_2|,~~~~\be=\max\{\ga_1,\ga_2\}.
\end{equation}
Fig.1 gives the mechanics analogue of fractional Maxwell's model,
where a triangle denotes a fractional element model.
\begin{figure}[ht]
\begin{center} \small
\includegraphics[height=40mm]{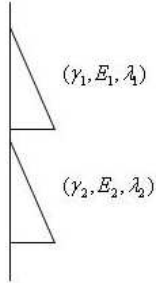}
\end{center}
\caption{Mechanics analogue of fractional Maxwell's
model. A triangle denotes a fractional element model.}\label{fig1}
\end{figure}

The continuous interest of fractional Maxwell's model is perhaps due
to the special known and unknown property of this model and due to
the rapid development of fractional calculus during the last fifty
years. Palade et. al. in \cite{Palade} derived fractional Maxwell's
model from the linearization of the objective equation and
discovered the anomalous stability behavior of the rest state in
three dimensions. Tan et. al. in \cite{Tan}\cite{Tan2} studied four
unsteady flows of a viscoelastic fluid with generalized Maxwell's
model between two infinite parallel plates. Vieru et. al. in
\cite{Fetecau_2} studied flow of a generalized Oldroyd-B fluid due
to a constantly accelerating plate, which includes generalized
Maxwell's model as a limiting case. Hayat et al. \cite{Hayat} also
discussed fractional Maxwell's model and studied three types of
unidirectional flows which were induced by general periodic
oscillations of a plate. Hern\'andez-Jim\'enez et. al. gave some
experimental results for oscillating flows with fractional Maxwell's
model \cite{Herna}, which encourage more studies. Yin and Zhu
\cite{YZ} studied the oscillating flow with fractional Maxwell's
model in an infinitely long pipe and found interesting results, for
example, the resonance peaks was discovered to be different with
those of ordinary Maxwell's model.

A lot of interests and studies were also given to the unidirectional
start-up pipe flows, which has a significant practical and
mathematical meaning. Zhu and Lu et. al. in \cite{Zhu_Lu} studied
characteristics of the velocity filed and the shear stress field for
an \emph{ordinary Maxwell's fluid} and discovered the oscillation
phenomenon. Fetecau in \cite{C_Fetecau} studied the analytic
solution for an \emph{ordinary} \emph{Oldroyd-B fluid} and several
limiting cases such as ordinary Maxwell's fluid; the velocity
profiles for the steady state are the same in all types of fluid
they studied. Further, Tong et. al. in \cite{OB} studied the exact
solution for the \emph{fractional Oldroyd-B model} in an annular
pipe by using Hankel-Laplace transform. Using similar but different
methods, Zhu and Yang et. al. in \cite{Zhu_Yang} studied the exact
solution and flow characteristics for the \emph{fractional element
model}\footnote{We mention that the fractional Oldroyd-B model in
the study of Tong et. al. does not include the fractional element
model as a limiting case.} with parameter $\be$, the most
fundamental model in all fractional derivative models. They found
oscillation phenomenon and solid-like behavior for certain $\beta$.

As far as we know, the characteristics of start-up pipe flow with
\emph{fractional Maxwell's model} have not been well-studied yet. In
this paper, we investigate basic characteristics of such flows
through studying the exact solution. The fluid is quiescent in the
beginning in an infinitely long pipe, and then it will be suddenly
started by a pressure gradient which remains constant after the
starting moment.

The start-up pipe flow with a dashpot model (i.e. Newtonian fluid)
is classical in fluid dynamics; with ordinary Maxwell's model the
motion would be very interesting because of the occurrence of
oscillations \cite{Zhu_Lu}; and with the fractional element model
solid-like behavior was discovered \cite{Zhu_Yang} as we already
mentioned above. One must be curious what the interesting phenomena
for fractional Maxwell's model are and whether we can summarize a
way to deduce the characteristic of a certain viscoelastic model if
just given the constitutive relation.

This paper is organized as follows. In section 2, we present the
governing equations and initial boundary conditions, and solve this
initial boundary value problem through the method of variable
separation and Heaviside's operational calculus. In section 3, we
discuss the flow characteristics and give some observations. In
section 4, we make the conclusion and propose a conjecture.

\section{Governing equation and exact solution}
We consider the start-up flow of a fractional Maxwell fluid in an
infinitely long pipe with the radius $a$. In the beginning, the
fluid in the pipe is at rest; then it is suddenly started by a
constant pressure gradient. Take the pipe axial direction as the
$z$-axis. We construct a column coordinate system $(r,\theta,z)$ and
let $V(r,\theta,z,t)$ denote the flow field.

Since the flow is axisymmetric, we assume $V$ only has the axial
component and does not depend on $\theta$, i.e.,
\begin{equation}\label{V}
V=u(r,t)e_z.
\end{equation}

The governing equations of the motion are given by the continuous
equation
\begin{equation}\label{conti}
\nabla\cdot V=0,
\end{equation}
as well as the momentum equation
\begin{equation}\label{momentum}
\rho\f{d V}{dt}=-\nabla p+\nabla\cdot \sigma,
\end{equation}
where $\f{d}{dt}$ is the material derivative, $\rho$ is the density
of the fluid, $p$ is the pressure field and $\sigma$ is the stress
tensor field.

The start-up flow considered is driven by the pressure gradient
field given by
\begin{equation}\label{press}
\f{\pa p}{\pa z}(t)=-G h(t),
\end{equation}
where $G$ is a constant and $h(t)$ is the Heaviside function,
defined by $h(t)=0,~t<0;~h(t)=1,~t\geq 0$.

The constitutive equation \eqref{constitutive} gives:
\begin{equation}\label{cons_rz}
\sigma_{rz}+\la^\al\f{\pa^\al}{\pa
t^\al}\sigma_{rz}=E\la^\be\f{\pa^{\be-1}}{\pa t^{\be-1}}\f{\pa
u}{\pa r}.
\end{equation}
The initial and boundary conditions are given by
\begin{equation}
u(a,t)=0,~~u(0,t)<\infty,~~~u(r,0)=\f{\pa u}{\pa
t}(r,0)=0,~~~~t>0,~0\leq r\leq a.
\end{equation}

Substituting \eqref{V} to equations \eqref{conti} and
\eqref{momentum}, and considering \eqref{press} and \eqref{cons_rz},
we find
\begin{equation}
\rho\Big(1+\la^\al\f{\pa^\al}{\pa t^\al}\Big)\f{\pa u}{\pa
t}=\Big(1+\la^\al\f{\pa^\al}{\pa t^\al}\Big)G h+E\la^\be
\f{\pa^{\be-1}}{\pa t^{\be-1}}\Big(\f{1}{r}\f{\pa u}{\pa r}+\f{\pa^2
u}{\pa r^2}\Big).
\end{equation}

Let the radius $a$ be the characteristic length, $\rho a^2/E\la$ be
the characteristic time, and $G a^2/E\la$ be the characteristic
velocity. By simple algebraic manipulations, we get the
dimensionless governing equation and the dimensionless
initial-boundary conditions\footnote{We still use notations $u,r,t$
but here they denote dimensionless quantities. In the followings we
only consider dimensionless quantities, so this change of notations
will not bring any confusions.}:
\begin{equation}\label{gov}
\f{\pa u}{\pa t}+\la^\al\f{\pa ^{\al+1}}{\pa
t^{\al+1}}u-\la^{\be-1}\f{\pa ^{\be-1}}{\pa
t^{\be-1}}\Big(\f{1}{r}\f{\pa u}{\pa r}+\f{\pa^2 u}{\pa
r^2}\Big)=1+\la^\al \f{t^{-\al}}{\Ga(1-\al)},
\end{equation}\begin{flushright}$0\leq r<1,~t>0,$\end{flushright}
and
\begin{subequations}\label{ib}
\begin{align}
&u(1,t)=0,~~u(0,t)<\infty,~~~t>0,\\
&u(r,0)=\f{\pa u}{\pa t}(r,t)=0,~~~0\leq r<1.
\end{align}
\end{subequations}

We use the method of variable separation and Heaviside operational
calculus solving equations \eqref{gov} and \eqref{ib}. Let
\begin{equation}
u(r,t)=v(r)T(t).
\end{equation}
Substituting this expression to the homogenous equation of equation
\eqref{gov}, we obtain
\begin{equation}
\f{1}{v}\Big(\f{1}{r}v'(r)+v''(r)\Big)=\f{T'(t)+\la^\al
T^{\al+1}(t)}{\la ^{\be-1}T^{\be-1}(t)}=-k^2,
\end{equation}
where $k$ is some appropriate constant to be determined. Solving the
eigenvalue problem:
\begin{subequations}
\begin{align}
&\f{1}{r}v'(r)+v''(r)+k^2 v=0,\\
&v(1)=0,~v(0)<\infty,
\end{align}
\end{subequations} we obtain the discrete eigenvalues
\begin{equation}
k_1<k_2<k_3<...,
\end{equation}
as well as the corresponding eigenfunctions:
\begin{equation}
v_m=J_0(k_m r),~~~~~m=1,2,...,
\end{equation}
where $k_m$ is the $m_{th}$ positive root of the zeroth Bessel
function.

The final solution is constructed by
\begin{equation}
u(r,t)=\sum_{m=1}^{\infty}A_m T_m(t) J_0(k_m r),
\end{equation}
where $A_m$, $m=1,2,3,...,$ are constants to be determined and
$T_m(t)$ are functions of $t$ to be determined. Substituting this
expression of $u(r,t)$ to equation \eqref{gov}, we find
\begin{equation}\label{1}
\sum_{m=1}^{\infty}A_m\big(T_m'(t)+\la^\al
T^{\al+1}(t)+k_m^2\la^{\be-1}T_m^{\be-1}(t)\big)J_0(k_m
r)=h(t)+\f{\la^\al t^{-\al}}{\Ga(1-\al)};
\end{equation}
since
 \begin{equation}\label{2}
   h(t)+\f{\la^\al t^{-\al}}{\Ga(1-\al)}=\Big(h(t)+\f{\la^\al
   t^{-\al}}{\Ga(1-\al)}\Big) \sum_{m=1}^{\infty}\f{2 J_0(k_m r)}{k_m J_1(k_m)},
   ~(0<r<1),
 \end{equation}
comparing the coefficients of the eigenfunctions appearing in
equations \eqref{1} and \eqref{2} we have
\begin{equation}
A_m=\f{2}{k_m J_1(k_m)},
\end{equation}
as well as
\begin{equation}\label{3}
T_m'(t)+\la^\al
T_m^{\al+1}(t)+k_m^2\la^{\be-1}T_m^{\be-1}(t)=h(t)+\f{\la^\al
t^{-\al}}{\Ga(1-\al)}.
\end{equation}

We solve equation \eqref{3} by applying Heaviside operational
calculus\footnote{Heaviside operational calculus is in fact
equivalent to Laplace transform method, but the former method is
more intuitive: The spectral parameter has a clear meaning.}. Let
$p=\f{d}{dt}$ and let $T_m(t)=Y h(t)$ where $Y$ is an operator to be
determined. Noting that
\begin{equation}
p^\la h(t)=\f{t^{-\al}}{\Ga(1-\al)},
\end{equation}
we have
\begin{equation}
Y=\f{1+\la^\al p^\al}{p+\la^\al
p^{\al+1}+k_m^2\la^{\be-1}p^{\be-1}}.
\end{equation}
As a result,
\begin{equation}
T_m(t)=\f{1+\la^\al p^\al}{p+\la^\al
p^{\al+1}+k_m^2\la^{\be-1}p^{\be-1}} h(t).
\end{equation}
By the definition of the Heaviside operator \cite{Courant}, it
yields
\begin{equation}
T_m(t)=\f{1}{2\pi \sqrt{-1}}\int_{L} \f{1+\la^\al z^\al}{z^2+\la^\al
z^{\al+2}+k_m^2\la^{\be-1}z^\be}e^{zt}dz.
\end{equation}
where $L$ is a contour in complex $z$-plane parallel with the
imaginary axis, and is determined by the requirement that there be
no singularities of the integrant on the right of $L$.

The final solution we construct is given by
\begin{equation}\label{final}
u(r,t)=\f{1}{2\pi \sqrt{-1}}\sum_{m=1}^{\infty} \f{2J_0(k_m r)}{k_m
J_1(k_m)} \int_{L} \f{1+\la^\al z^\al}{z^2+\la^\al
z^{\al+2}+k_m^2\la^{\be-1}z^\be}e^{zt}dz.
\end{equation}
Substituting this equation to equation \eqref{cons_rz} and assuming
the shear stress field is zero at $t=0$, we obtain
\begin{equation}\label{shear}
\sigma_{rz}=\f{E\la^{\be}}{2\pi \sqrt{-1}}\sum_{m=1}^{\infty}
\f{-2J_1(k_m r)}{J_1(k_m)} \int_{L} \f{z^{\be-1}}{z^2+\la^\al
z^{\al+2}+k_m^2\la^{\be-1}z^\be}e^{zt}dz.
\end{equation}

In the case of $\al=0$, i.e., the case of the fractional element
model, solution \eqref{final} reduces to
\begin{equation}\label{final_1}
u(r,t)=\f{1}{2\pi \sqrt{-1}}\sum_{m=1}^{\infty} \f{2J_0(k_m r)}{k_m
J_1(k_m)} \int_{L} \f{2}{2z^2+k_m^2\la^{\be-1}z^\be}e^{zt}dz.
\end{equation}
By an inverse formula given in \cite{Pod}[p.p. 271-273], we can
simplify expression \eqref{final_1}:
\begin{equation}\label{final_ele}
u(r,t)=\sum_{m=1}^{\infty} \f{2J_0(k_m r)}{k_m J_1(k_m)} t
E_{2-\be,2}(-k_m^2 \la^{\be-1}t^{2-\be}/2),
\end{equation}
where $E_{2-\be,2}$ is the Mittag-Leffler function. Note that the
Newtonian fluid corresponds to the particular case of
$(\al=0,~\be=1)$. Substituting $\be=1$ to equation
\eqref{final_ele}, we obtain
\begin{equation}
u(r,t)=\sum_{m=1}^{\infty} \f{4J_0(k_m r)}{k_m^3 J_1(k_m)}
\big(1-e^{-k_m^2 t/2}\big)=\f{1}{2}(1-r^2)-\sum_{m=1}^{\infty}
\f{4J_0(k_m r)}{k_m^3 J_1(k_m)} e^{-k_m^2 t/2},
\end{equation}
which is the classical dimensionless solution.

In the case of $\al=\be=1$, i.e., the case of ordinary Maxwell's
model, our solution agrees with the solution obtained by Zhu et. al.
\cite{Zhu_Lu}.

\section{Results and discussions}
Due to the simplicity of the fractional element model, which also
plays the fundamental role of constructing different fractional
models, we first recall some results in \cite{Zhu_Yang} of start-up
pipe flow for the fractional element model (Scott Blair's model).
Without lost of generality, we take $\la=1$ in equation
\eqref{final_ele}. As Fig.2 shows, for the case
$$0<\be<1,$$ oscillations occur just after the fluid is started; the
smaller the parameter $\be$ is, the stronger the elasticity is. And
as $t$ goes to infinity, the center velocity tends to be $0$. In fact,
recall the asymptotic formula for $E_{2-\be,2}$ \cite{Pod}:
\begin{equation}
E_{2-\be,2}(-z)=
\f{1}{\Ga(\be)}\f{1}{z}+O(z^{-2}),~~~~~~z\rightarrow +\infty,
\end{equation}
We know from this formula that the series expression
\eqref{final_ele} is uniformly convergent at least for any fixed
$r$. Let $t\rightarrow \infty$; we obtain that
\begin{equation}
\lim_{t\rightarrow \infty} u(r,t)=0,~~~~~~~0\leq r\leq 1.
\end{equation}
The only exception is the case of $\be=1$, i.e. the case of
Newtonian fluid. No oscillations would occur and as $t$ goes to
infinity, the center velocity will tend to be a steady constant
$0.5$. In this classical case,
\begin{equation}
\lim_{t\rightarrow \infty} u(r,t)=\f{1}{2}(1-r^2),~~~~0\leq r\leq 1.
\end{equation}
\begin{figure}[ht]
\begin{center} \small
\includegraphics[width=0.95\textwidth]{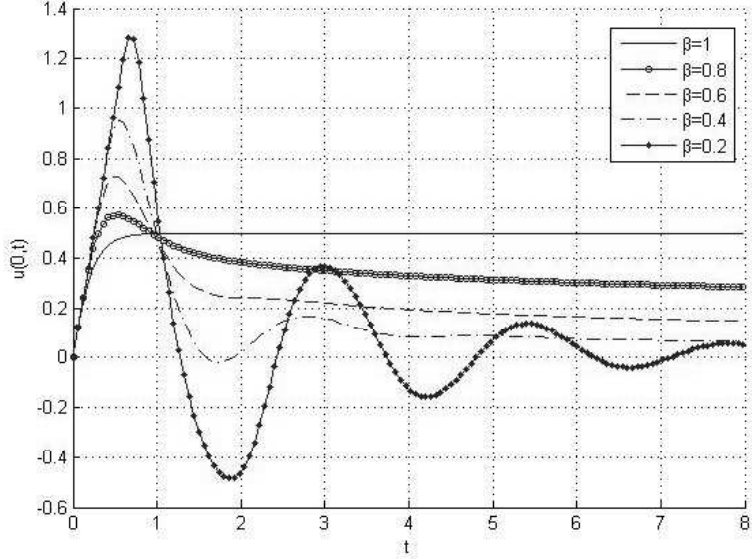}
\end{center}
\caption{Center velocity with respect to $t$ for the fractional
element model.}\label{fig2}
\end{figure}
Fig. 3a) and 3b) shows the velocity profiles of different $t$, with
$\be=0.4$ and $\be=1$ respectively, which gives intuitive pictures
in mind.
\begin{figure}[ht]
\begin{center} \small
$$
\begin{array}{cc}
\includegraphics[width=0.48\textwidth]{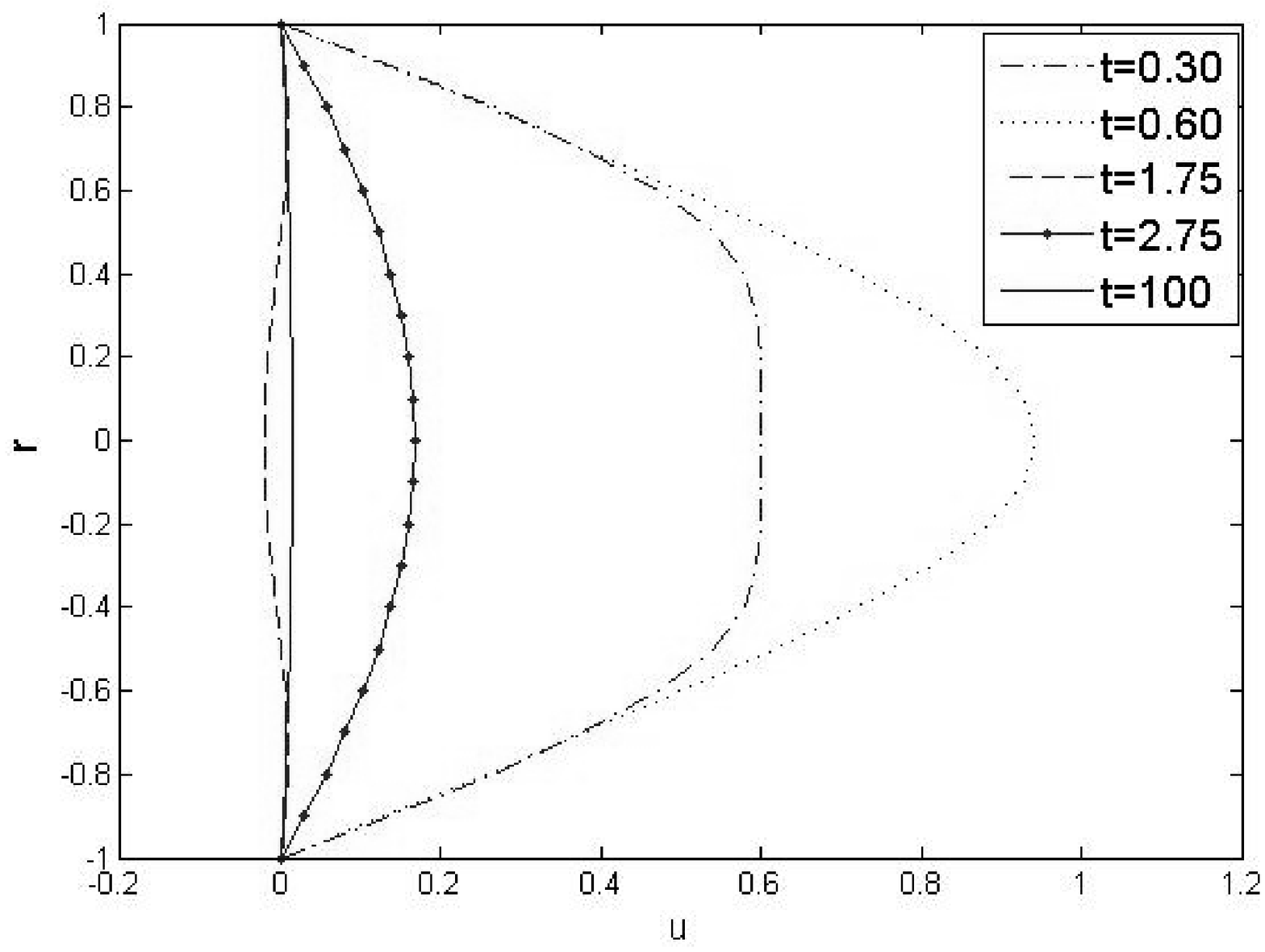} &
\includegraphics[width=0.45\textwidth]{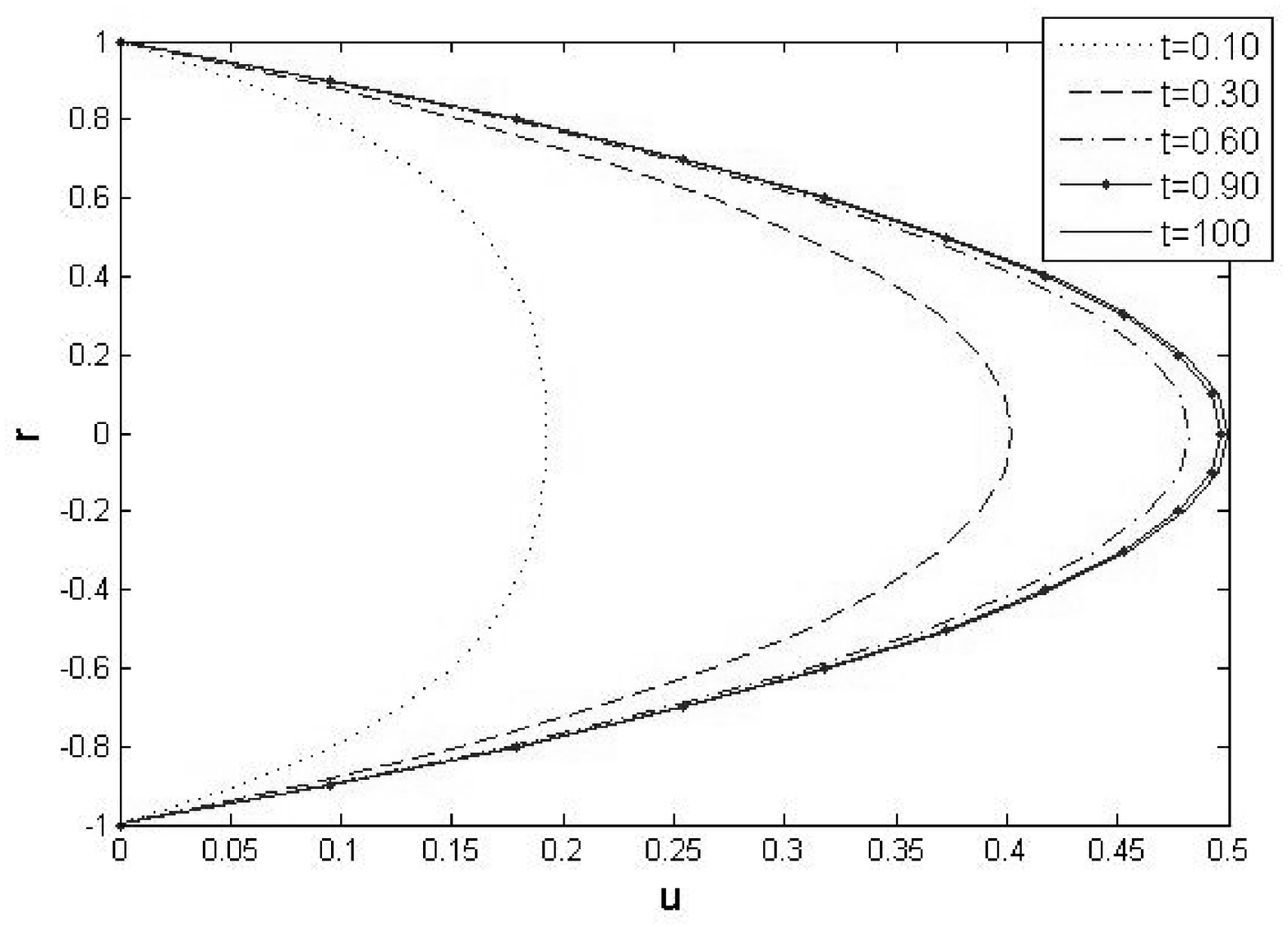} \\
\mbox{a.}~~ \beta=0.4 & \mbox{b.}~~ \beta=1 \\
\end{array}
$$
\end{center}
\caption{Velocity profiles at different t for the fractional element
model.}\label{fig3}
\end{figure}

Based on these discussions of of the fractional element model, the
mechanics analogue of fractional Maxwell's model (see Fig.1) would
help us do further investigates. This idea will be directly applied
in the following discussions. Still we will take $\la=1$ without
loss of generality. And we discuss the start-up flows with different
$\al$ and $\be$ through studying equation \eqref{final}.

Fig.4 gives the curves of the center velocity with respect to $t$
when $\al=0.6$. It can be seen that for $0.6\leq\be<1$, the center
velocity of the pipe will tend to be $0$, which means the
corresponding fractional Maxwell's model will finally represents
solid-like property; for $\be=1$, the center velocity will tend a
constant $0.25$ and the fractional Maxwell model finally represents
fluid-like property. Furthermore, the smaller $\be$ is, i.e., the
stronger the elasticity of the larger order fractional element is,
the stronger the oscillating phenomenon is.
\begin{figure}[ht]
\begin{center} \small
\includegraphics[width=0.95\textwidth]{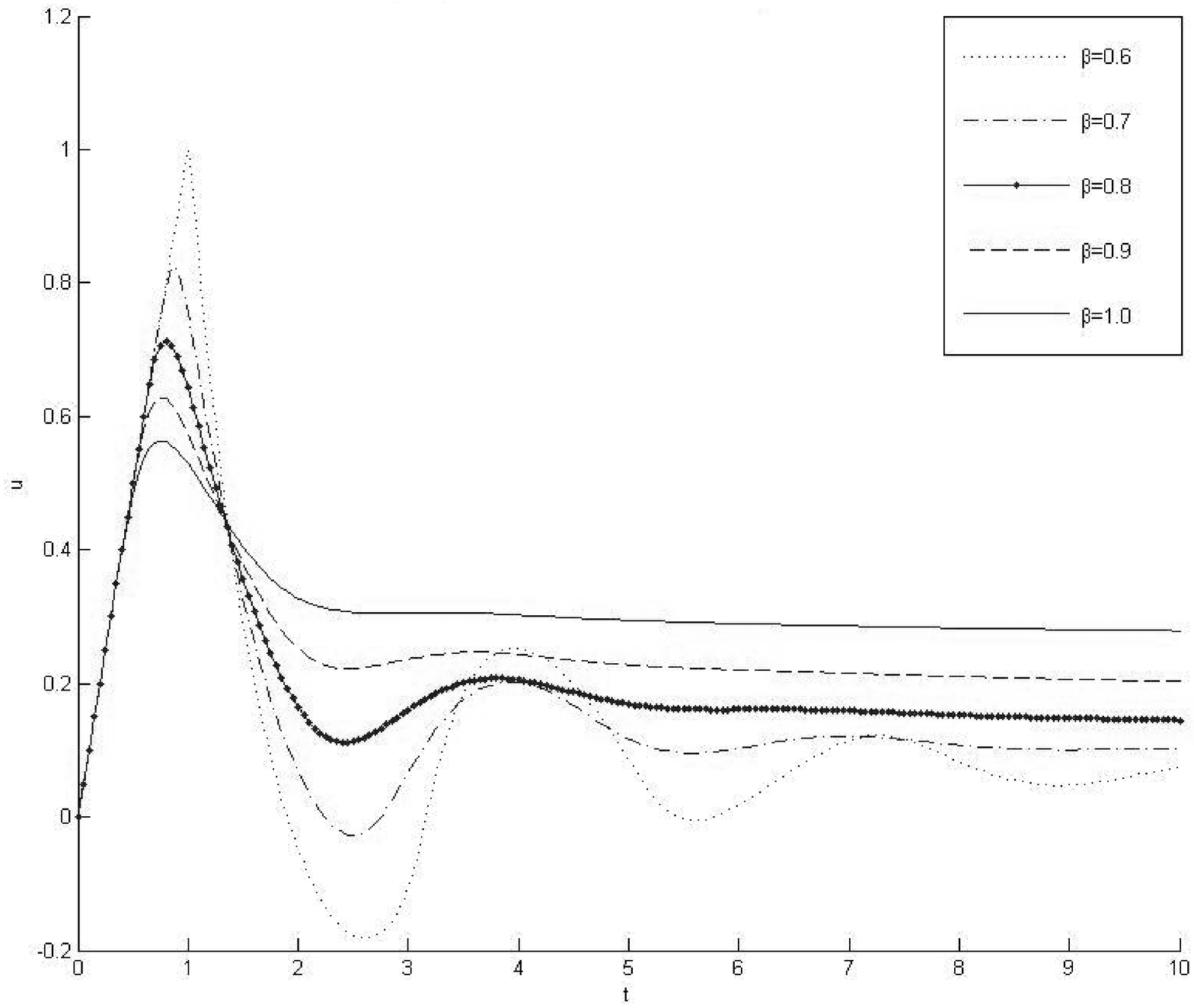}
\end{center}
\caption{Center velocity with respect to $t$ for the fractional
Maxwell model in the case of $\al=0.6$.}\label{fig4}
\end{figure}

Fig.5 is similar with Fig.4 but the phenomenon is more clear. It
gives the relation curve of center velocity with respect to $t$ with
different $\al$, when $\be=0.6$. It can be seen that, as $t$ goes to
infinity, the center velocity of the pipe tends to be $0$, which
indicates the corresponding fractional Maxwell's model represents
solid-like property. When $\al$ increases, i.e. the difference
between the orders of the two fractional elements increases, or say
the elasticity of the smaller order fractional element strengthens,
the oscillating phenomenon becomes stronger, which helps the fluid
to be at rest, although in short time it accelerates the fluid more.
\begin{figure}[ht]
\begin{center} \small
\includegraphics[width=0.95\textwidth]{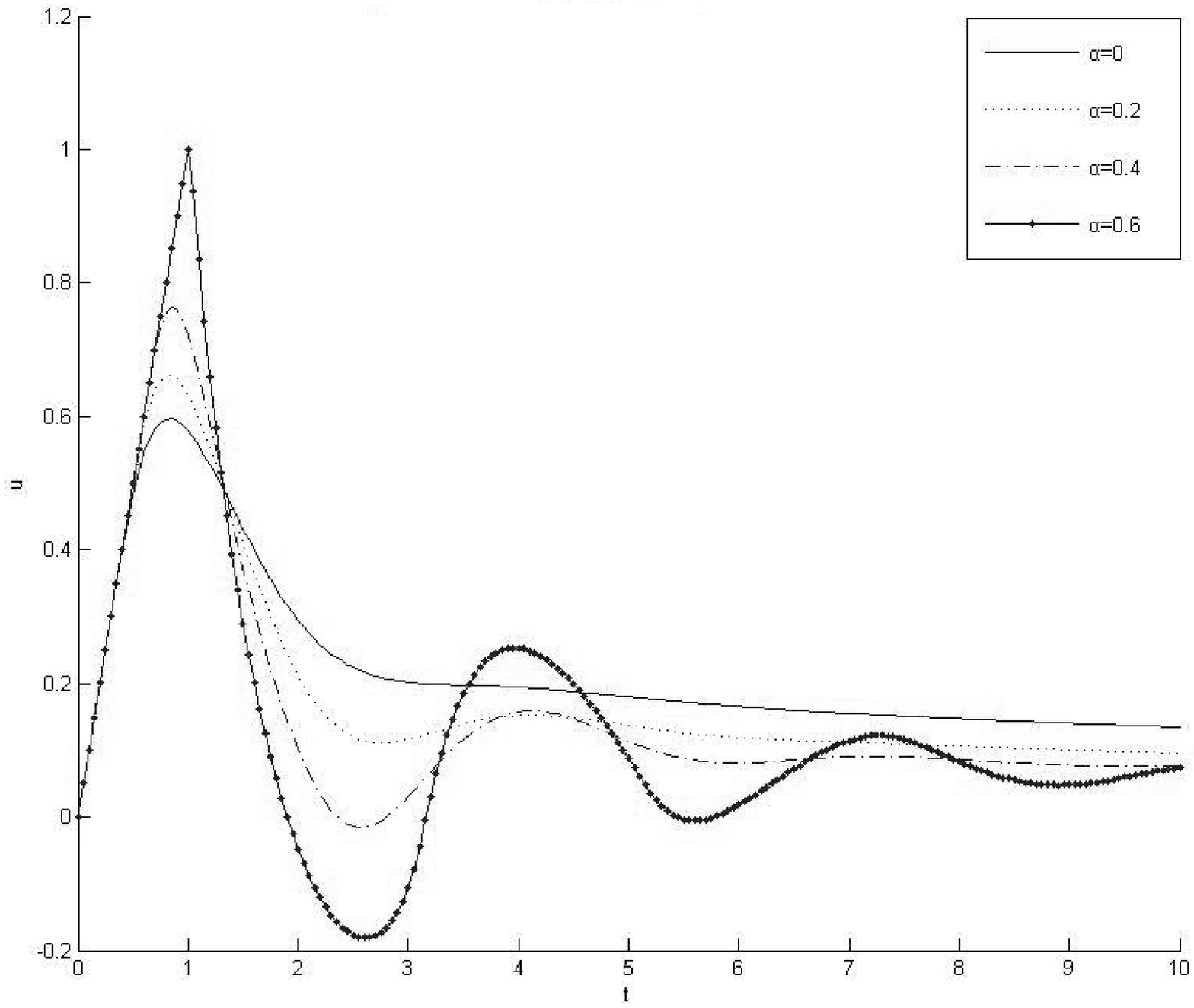}
\end{center}
\caption{Center velocity with respect to $t$ for the fractional
Maxwell model in the case of $\be=0.6$.}\label{fig5}
\end{figure}

Fig.6 shows the relation curve of the center velocity with respect
to $t$ in the case of $\be=1$. We see that if $0<\al\leq 1$, the
center velocity will tend to a constant $0.25$ as $t$ goes to
infinity; if $\al=0$, i.e., the case of Newtonian fluid, the center
velocity will tend to $0.5$. In both cases, fractional Maxwell's
fluid will represent fluid-like property for large $t$. And when
$\al\neq 0$, oscillation also occurs which means the fluid also has
elasticity, however, since $\be=1$, the fluid-like property will
lead the way.
\begin{figure}[ht]
\begin{center} \small
\includegraphics[width=0.95\textwidth]{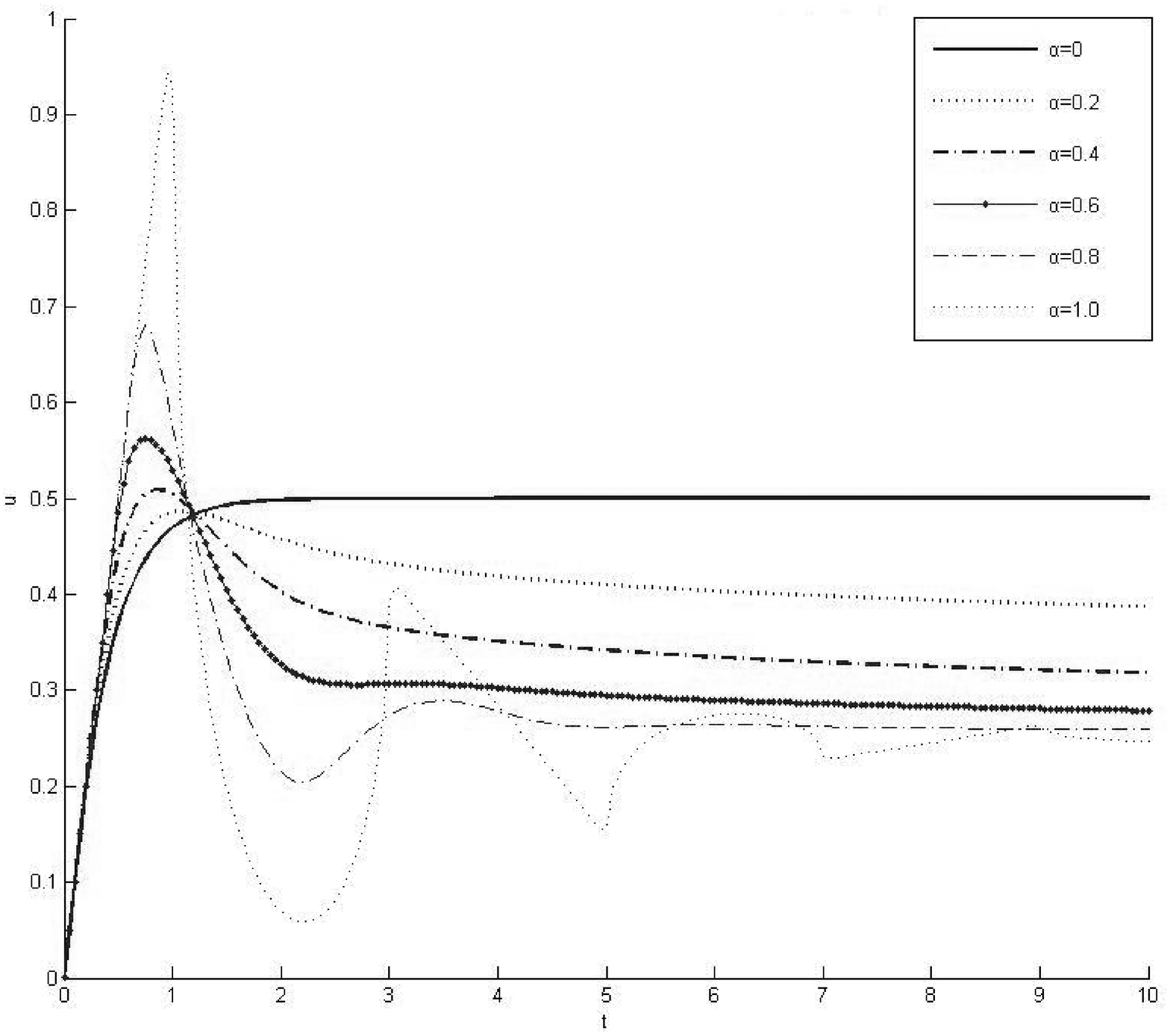}
\end{center}
\caption{Center velocity with respect to $t$ for the fractional
Maxwell model in the case of $\be=1$.}\label{fig6}
\end{figure}

It deserves to point out that these results partly agree with
Friedrich's \cite{Friedrich}, who first pointed out that the
fractional Maxwell model represents solid-like property as long as
$\be<1$.

An intuitive way to deduce whether a fractional model would exhibit
solid-like or fluid-like property is to consider the fractance of
the mechanics analogue of the model. According to the analysis of
the fractal construction of these models \cite{Pod}\cite{ZHY}, we
can indeed deduce solid-like property without calculations: A
fractional model (which itself is not a string) presents solid-like
behavior if and only if there exists a spring path from one side to
the other in any spring-dashpot fractance of its mechanics
analogues.

\section{Conclusions}
An exact solution of the start-up pipe flow with fractional
Maxwell's model is obtained and the flow characteristics are
discussed.

In the case of  $\beta\neq 1$, the motion of fractional Maxwell's
fluid in a pipe tends to be at rest as time goes to infinity;
otherwise if $\beta=1$ the flow will have a parabolic-like profile
as $t$ goes to $\infty$.

Indeed, for a Scott-Blair's model (fractional element model) with
parameter $\beta$, as long as $\beta<1$, the model presents a
solid-like property. This was showed by Zhu and Yang et. al. in
\cite{Zhu_Yang} in the study of start-up pipe flow.

For a fractional Maxwell model, i.e., two fractional element models
in a serial connection, the fluid will present a solid-like property
as long as the two fractional elements both present solid-like
behaviors. On the other hand, if one of the fractional elements in
serial is Newtonian, even the other one is a spring, the serial
compound will be of fluid-like property and the flow will keep a
stationary velocity profile after infinitely long time. Our result
agrees with the results in \cite{Friedrich}, but from different
points of views.

Moreover, the stronger the elasticity of any of the fractional
element models in the serial connection is, the stronger the
oscillation of the corresponding fractional Maxwell model is.

From these results for the case of fractional Maxwell's model, we
conjecture that a viscoelastic model presents a solid-like behavior
if for any spring-dashpot fractance of its mechanics analogues,
there exists a \emph{spring path}(a path with only springs) from one
side to the other. This conjecture, if proved, will generalize
Friedrich's result in \cite{Friedrich}.

\paragraph{Acknowledgements}
We are very grateful to the referee for suggestions and constructive comments.

\bibliographystyle{amsplain}

\begin{thebibliography}{10}
\bibitem{YZ}
Y. Yin, K.-Q. Zhu, {\em Oscillating flow of a viscoelastic fluid in
a pipe with the fractional Maxwell model}, Applied Mathematics and
Computation, 173 (2006) 231-242.

\bibitem{ZHY}
K.-Q. Zhu, K.-X. Hu, D. Yang, {\em Analysis of fractional element of
viscoelastic fluids using Heaviside operational calculus}, New
trends in fluid mechanics research, Edited by F.-G. Zhuang and J.-C.
Li, Tsinghua University Press, Springer, (2007) 506-509.

\bibitem{Tan}
W. Tan, W. Pan, M. Xu, {\em A note on unsteady flows of a
viscoelastic fluid with the fractional Maxwell model between two
parallel plates}, Int. J. Non-Linear Mech. 38 (2003) 645-650.

\bibitem{Herna}
A. Hern\'andez-Jim\'enez, J. Hern\'andez-Santiago, A.
Macias-Garc\'ia, J. Sanchez-Gonz\'alez, {\em Relaxation modulus in
PMMA and PTFE fitting by fractional Maxwell model}, Polym. Test. 21
(2002) 325-331.

\bibitem{Hayat}
T. Hayat, S. Nadeem, S. Asghar, {\em Periodic
unidirectional flows of a viscoelastic fluid with the fractional
Maxwell model}, Appl. Math. Comput. 151 (2004) 153-161.

\bibitem{G. W. Scott Blair}
G. W. Scott Blair,  {\em The role of Psychophysics in rheology},
Journal of Colloid Science, 2 (1947) 21-32.

\bibitem{Scott2}
G. W. Scott Blair, {\em Psychoreology: link between the past and the present}, Journal
of Texture Studies, Vol.5 (1974) 3-12.

\bibitem{Heymans}
N. Heymans, J. C. Bauwens, {\em Fractal rheological models and fractional
differential equations for viscoelastic behavior}, Rheol. Acta 33 (1994) 210-219.

\bibitem{Tan2} W. Tan, M. Xu, {\em Plane surface suddenly set in motion in a viscoelastic fluid with
fractional Maxwell model}, ACTA MECHANICA SINICA, 18 (2002).

\bibitem{Pod}
I. Podlubny, {\em Fractional Differential Equations}, London: Academic
Press, (1999).

\bibitem{Courant} R. Courant, D. Hilbert, {\em Methods
of mathematical physics}, v.1, New York: Interscience Publishers,
Inc. (1962).

\bibitem{Friedrich} C. Friedrich, {\em Relaxation and
retardation functions of the Maxwell model with fractional
derivatives}, Rheol. Acta 30, (1991) 151-158.

\bibitem{Zhu_Yang}
K.-Q. Zhu, D. Yang, K.-X. Hu, {\em Fractional element of viscoelatic
fluids and start-up flow in a pipe}, Chinese Quaterly of Mechanics,
v. 28, No. 4,  (2007). (in Chinese).

\bibitem{Zhu_Lu}
K.-Q. Zhu, Y.-J. Lu, P. P. Shen, J. L. Wang, {\em A study of
start-up pipe flow of Maxwell fluid}, Acta Mech. Sin., v. 35, 218,
(2003). (in Chinese)

\bibitem{Palade}
L. I. Palade, P. Attane, R. R. Huilgol, B. Mena, {\em Anomalous
stability behavior of a properly invariant constitutive equation
which generalises fractional derivative models}, International
Journal of Engineering Sciences, 37 (1999), 315-329.

\bibitem{OB}
D. Tong, R. Wang, H. Yang, {\em Exact solutions for the flow of
non-Newtonian fluid with fractional derivative in an annular pipe},
Science in China ser: G Physics, Mechanics and Astronomy, Vol. 48,
No. 4, (2005) 485-495.

\bibitem{C_Fetecau}
C. Fetecau, {\em Analytical solutions for non-Newtonian fluid flows
in pipe-like domains}, Non-linear Mechanics 39 (2004), 225-231.

\bibitem{Fetecau_2}
D. Vieru, Corina Fetecau, C. Fetecau, {\em Flow of a generalized
Oldroyd-B fluid due to a constantly accelerating plate}, Applied
Mathematics and Computation 201 (2008) 834-842.

\bibitem{Fetecau_3}
M. Khan, S. Hyder Ali, C. Fetecau, Haitao Qi, {\em Decay of
potential vortex for a viscoelastic fluid with fractional Maxwell
model}, Applied Mathematical Modelling 33 (2009) 2526-2533.

\bibitem{Fetecau_4}
D. Vieru, Corina Fetecau, M. Athar, Constantin Fetecau, {\em Flow of
a generalized Maxwell fluid induced by a constantly accelerating
plate between two side walls}, Z. angew. Math. Phys. 60 (2009)
334-343.
\end{thebibliography}

\end{document}